\documentclass{iopart}
\usepackage{graphicx,epsf}
\begin{document}
\title{Quadrupole correlations and inertial properties of rotating nuclei}
\author{D Almehed\dag, F D\"{o}nau\ddag and  R G Nazmitdinov\S\P}
\address{\dag\ Department of Physics, UMIST, P.O. Box 88, Manchester M60 1QD, 
United Kingdom}
\ead{D.Almehed@umist.ac.uk}
\address{\ddag\ IKH, FZ Rossendorf, PF 510119, D-01314 Dresden, Germany}
\address{\S Departament de F{\'\i}sica, 
Universitat de les Illes Balears, E-07122 Palma de Mallorca, Spain}
\address{\P Bogoliubov Laboratory of Theoretical Physics,
Joint Institute for Nuclear Research, 141980 Dubna, Russia}

\date{\today}

\begin{abstract}
The contribution of quantum shape fluctuations to inertial properties
of rotating nuclei has been analysed for QQ--nuclear interaction
using the random phase approximation (RPA). The different 
recipes to treat the cranking mean field plus RPA problem are considered.
The effects of the $\Delta N=2$ quadrupole matrix elements and
the role of the volume conservation condition are discussed. 
\end{abstract}

\submitto{\JPG}
\maketitle

\section{Introduction}

Response of a many-body system to external fields
provides various information about intrinsic dynamics of 
the system. One of the efficient approaches to get a deep insight 
into a nuclear structure is to study a response of a nucleus to an 
external Coriolis field. In particular, we can trace the interplay 
between single-particle and collective degrees of freedom by studying 
a rotational dependence of the kinematical and dynamical moments of  inertia. 
In its turn, these quantities are benchmarks for nuclear microscopic models,
that allow to understand  main features of a nuclear field.
For example, the importance of pairing correlations,
introduced in nuclear physics by
Bogoliubov \cite{Bog} and Bohr, Mottelson and Pines \cite{BMP} 
in an analogy with the correlations in superconductors,
had been recognised first in the
description of nuclear inertial properties \cite{Be59}.

The kinematical moment of inertia is described quite well
within the cranking approach, for instance, in the relativistic 
Hartree-Bogoliubov and non-relativistic density-dependent 
Hartree-Fock-Bogoliubov theory with Gogny forces (see \cite{AK00} and 
references therein). However, there is
a systematic discrepancies between those theories and experiment
with regard of the dynamical moment of inertia.
It is quite desirable to understand the main source of discrepancies between
experiment and theory, since the above approaches represent state of art of
nuclear structure studies.

It is well understood that the two-body correlations are important
for a correct description of the moments of inertia (see discussions in
\cite{AK00}). One of the systematic ways to include these correlations 
is the random phase approximation (RPA) based on a self-consistent solution
 of the cranking mean field equations \cite{Th60,Th61}. 
 This enables one to also restore symmetries broken on the mean field level 
 (see textbooks \cite{RS80,BR86}). 
Practical realisation of these aspects has been done recently in 
a cranking harmonic oscillator model with a residual 
quadrupole--quadrupole interaction \cite{NA02}.
In particular, it was shown that the Thouless-Valatin moment 
of inertia \cite{TV62} (which, in general, 
contains Belyaev \cite{Be59}, Migdal \cite{Mi59} and other terms) 
calculated in the RPA is equivalent to the dynamical moment of inertia 
calculated in the mean field approximation
\begin{equation}
{\cal J}_{TV} \equiv {\cal J}_x^{(2)}=
\frac{d}{d\Omega}\langle\Omega|{\hat J}_x|\Omega\rangle
\equiv -\frac{d^2E_{MF}}{d{\Omega}^2}.
\end{equation} 
Here  $\Omega$ denotes the rotational frequency, 
$|\Omega\rangle$ is a self-consistent solution of the mean field equations
and $E_{MF}$ is a total mean field energy in the rotating frame.

Quantum oscillations around the mean field solutions provide
the additional contribution to the total energy. 
How such a quantal effect can be obtained within the RPA  
for the case of rotating systems is generally described in~\cite{Ma77,KN86}.
However, the practical application of the RPA in rotating deformed
nuclei is a demanding numerical task. Therefore,    
the influence of quantum fluctuations on the moment of inertia 
is till now scarcely studied. With regard of the easier case of 
pairing fluctuations more extensive investigations were performed 
(c.f.~\cite{SG89,AF01}). The first attempt to incorporate the 
pairing and quadrupole oscillations \cite{EM80} was done in a 
restricted configuration space and 
the self-consistency between the residual interaction and 
the mean field was not taken into account.
A self-consistent analysis carried out in a simplified model \cite{NA02} 
demonstrated that the quantum correlations 
modify the Thouless-Valatin moment of inertia. Therefore, there is
a strong motivation to understand the role of quantum correlations
in realistic calculations for inertial properties.

The aim of the present paper is to study 
exclusively the influence of shape vibrations
on the angular momentum and on the moment of inertia. 
Starting from the self-consistent  
cranking mean field calculation the total binding energy as a function of 
the rotational frequency has to be calculated up to RPA order,
i.e. with inclusion of the RPA correlation energy.
Our calculations are based on separable QQ-forces as an effective interaction.
The mean field part of such calculations could be carried out 
with more advanced effective interactions like e.g. Skyrme or Gogny type forces.

However, the RPA correlation energy is not feasible 
for non-separable forces because of the 
too large configuration space needed for the cranking model.
In addition, the problem of spurious solutions in the RPA calculations
for the state of art effective interactions is silently avoided in the 
literature even for a zero rotational frequency. 
In other words, again the question arises about a consistency between 
a mean field calculations and the RPA contribution for those type of forces.
On the other hand, the  QQ-forces are the most important interaction 
term for inducing nuclear deformation and rotational excitations.
The mean field plus RPA calculations with the QQ-forces
based an a realistic cranking mean field 
potential with a spin-orbit coupling are by no means trivial. 
As we shall describe (see below) such calculations can be executed in 
different ways which  enable us study  specific properties of these forces 
and their effects in more detail.

We will consider the following versions of the QQ-model:\\
(i) the QQ-forces  with a full N-mixing of oscillator shells 
($\Delta N=0,\pm 2$);\\
(ii) the QQ-forces without N-mixing ($\Delta N=0$)  which is the commonly 
     used a QQ-model (e.g.~\cite{RS80}); \\
(iii)the double-stretched QQ-forces with the volume conservation constraint 
     ~\cite{Ab85,SK89}.\\
 By comparing results of the various treatments (i-iii) 
 provides new insights in the nature of both the QQ-forces and their influence 
 onto the rotational properties. Because of the schematic character of the
 QQ-forces our investigations are merely an exploratory approach to the 
 influence of quantum shape fluctuations upon rotational properties rather  
 than a quantitative description of experimental
 data. The latter would require more extensive numerical efforts that are
 not intended in this paper. 
  
The paper is organised as follows. In Sec.~\ref{model} we introduce our 
mean field and RPA models. The influence of the volume conservation 
constraint on the RPA correlations is discussed in Sec.~\ref{volume}. 
This is followed by a summary and discussion 
in~Sec.~\ref{summary}.

\section{ The Model}
\label{model}
Our mean field plus RPA calculations are based on the following Hamiltonian:
\begin{equation}
  \label{eq:QQ1}
  \hat{H} = \sum_k{\epsilon _k} \hat{c}_k^\dagger \hat{c}_{k} -
  \frac{\kappa}{2} \sum_{m=0, \pm 1, \pm 2}\hat{Q}_m^2  =H_{\rm sph} + H_{QQ}.
\end{equation}
Here $\epsilon _k$ are the single-particle energies of the 
spherical oscillator Hamiltonian $\hat{h}_{\rm sph}$ 
\begin{equation}
  \label{eq:Nilsson1}
  \hat{h}_{\rm sph} = \frac{\hat{p}^2}{2M}+
    \frac{M}{2} \omega_0^2  \hat{r}^2 .
\end{equation}
The  operators  $\hat{c}^\dagger_k$ ($\hat{c}_k$) are creation 
(annihilation) operators
with the suffix $k$  labeling a set of quantum numbers.
For the sake of convenience we have chosen in equation~(\ref{eq:QQ1})
a quadratic form of the QQ-forces by taking a set of Hermitean quadrupole 
operators $\hat{Q}_m$ built up from $\hat{r}^2 \hat{Y}_{20}$ and the linear 
combinations $\hat{r}^2 ( \hat{Y}_{2m} \pm \hat{Y}_{2-m})$ for $m= (1,2)$.

The operators $\hat{Q}_m$ read as
\begin{eqnarray}
  \label{eq:Q1}
  \hat{Q}_0  
  &=&   \sqrt{\varphi_0}\sum_{kl} q_{0,kl} \left( \hat{c}_k^\dagger \hat{c}_l 
  + \hat{c}_{\bar{k}}^\dagger
    \hat{c}_{\bar{l}} \right), \\
  \label{eq:Q2}
  \hat{Q}_{\pm 1}
   &=& \sqrt{\varphi_{\pm 1}}\sum_{kl} q_{\pm 1,kl} 
   \left( \hat{c}_k^\dagger \hat{c}_{\bar{l}}
    \pm \hat{c}_{\bar{k}}^\dagger \hat{c}_l \right), \\
  \label{eq:Q3}
  \hat{Q}_{\pm 2} &=& \sqrt{\varphi_{\pm 2}}\sum_{kl} q_{\pm 2,kl} 
  \left( \hat{c}_k^\dagger \hat{c}_l \pm
    \hat{c}_{\bar{k}}^\dagger \hat{c}_{\bar{l}} \right)
\end{eqnarray}
with
\begin{equation}
  \label{eq:QQpRPA3}
  \varphi_m = \left\{
    \begin{array}{ll}
      1 & m = 0, -1, 2\\
      -1 & m = 1, -2 
    \end{array}
    \right.
\end{equation}
where the index $k(l)$ is labeling a complete set of quantum numbers
in order to form the matrix elements $q_{m,kl}\,=\langle k|\hat{Q}_m|l\rangle$.

The index $\bar k$ refers to the time conjugated state. 
The sums in equations~(\ref{eq:Q1}--\ref{eq:Q3}) split in a proton and 
neutron part.
We consider the isoscalar quadrupole interaction only, since 
isovector terms can be treated in a similar way.
Hereafter, any isospin index is omitted.

The Hamiltonian $\hat H$, equation~(\ref{eq:QQ1}) is, of course, formally the 
same expression for all the versions  (i-iii) of the QQ-model described 
in the introduction. What is different concerns the particular treatment of 
the Q-operator (N-mixing or not and stretching transformation), 
the adjustment of the spherical oscillator frequency, $\omega_0$, to account 
for the volume conservation condition and the adaption of the strength 
parameter $\kappa$. This is because the rotational invariance of the total 
Hamiltonian is required for a clean treatment of the angular momentum 
conservation within the RPA \cite{Ma77,KN86}.

We consider the Routhian 
\begin{equation}
  \label{eq:Routh}
  \hat{H}' \,=\, \hat H\, -\,  \Omega \hat{J}_z
\end{equation}
to describe the rotational properties of the system.
Here $\hat{J}_z$ is
the angular momentum component about the z-axis. Note 
that the cranking term in $\hat H'$
does not change the exact wave functions but only fixes 
the angular momentum about the quantisation axis (z).

The relevant mean field part of the Routhian $H'$ is originated from
the Hamiltonian $H$ in equation (\ref{eq:QQ1}). Writing the latter term  
in the standard form of a deformed potential (c.f. \cite{NR95}), 
the mean field part of $H'$ reads as 
\begin{equation}
  \label{eq:QQ2}
  \hat{H}_{\rm MF}' = \hat{H}_{\rm sph}  -\Omega \hat{J}_z
  - \sqrt{\frac{5}{4\pi}} \hbar \omega_0 \beta
  \left( \hat{Q}_0 \cos \gamma -\hat{Q}_2 \sin \gamma \right)
\end{equation}
where $\beta$ and $\gamma$ are the deformation parameters.  
The self-consistent solution $\Phi^\omega \equiv |\Omega\rangle$ 
shortly denoted as $\,\,\rangle$
corresponds to an energy minimum of the 
energy surface $E'(\beta,\gamma)\,=\,\langle\hat H'\rangle$.

We are aiming at the contributions of zero point quantum corrections to the 
moment of inertia stemming from the RPA vibrations about 
the above mean field solution.   
Accordingly, the quasi boson approximation (QBA) is applied to the 
Routhian $\hat H'$, equation~(\ref{eq:Routh}), which is rewritten in the
form
\begin{equation}
  \label{eq:QQ4}
  \hat{H}' = \hat{H}'_{MF} + \frac{\kappa}{2} \sum_m \left<\hat{Q}_m \right>^2
  -\hat{H}_{\rm res}
\end{equation}
Here, the third term is a separable residual interaction
\begin{equation}
  \label{eq:QQint}
  \hat{H}_{\rm res} = \frac{\kappa}{2} \sum_m \left(\hat{Q}_m -\left< \hat{Q}_m
    \right>\right)^2
\end{equation}
neglected in the mean field calculations. 

Using the standard notation we refer to
particle states (unoccupied single-particle orbitals) 
by subscript $p$ and to 
hole states (below the Fermi level) by $h$.
Defining the boson-like operators 
\begin{equation}
  \label{eq:boson1}
  \hat{b}^\dagger_{ph} \equiv \hat{c}^\dagger_p \hat{c}_h \,\,\,\,,
  \,\,\,\,\hat{b}_{ph} \equiv \hat{c}^\dagger_h \hat{c}_p 
\end{equation}
the QBA means to treat the above
operators~(\ref{eq:boson1}) as exact bosons obeying the commutation
relations
\begin{equation}
  \label{eq:boson2}
  \left[ \hat{b}_\mu , \hat{b}_\nu \right] = \left[ \hat{b}^\dagger_\mu ,
        \hat{b}^\dagger_\nu
  \right] = 0 , \qquad \left[ \hat{b}_\mu , \hat{b}^\dagger_\nu \right] =
  \delta_{\mu \nu} 
\end{equation}
where the double index $ph$ runs over all particle-hole pairs 
and is labelled by $\mu$
or $\nu$, respectively. 
In this approximation any single-particle operator $F$ can be expressed as
\begin{equation}
\label{def}
{F} = \langle {F} \rangle + F^{(1)}+F^{(2)}
\end{equation}
where the second and third terms are linear and bilinear order terms in 
the boson expansion, respectively.

The final RPA Hamiltonian reads
\begin{equation}
  \label{eq:QQpRPA1}
  \hat{H}'_{\rm RPA} = \sum_\mu  E_\mu 
  \hat{b}^\dagger_\mu \hat{b}_\mu - \frac{\kappa}{2}
  \sum_m \left( \hat{Q}_m - \left< \hat{Q}_m \right> \right)^2
\end{equation}
where $E_\mu=e'_p-e'_h$ is the energy of a particle-hole excitation
with respect to the mean field part ${\hat H}_{\rm MF}'$, 
equation~(\ref{eq:QQ2}). We remind that in the QBA one includes 
all second order terms into the boson Hamiltonian such 
that $(F-<F>)^2=F^{(1)}F^{(1)}$.

The RPA Hamiltonian is diagonalised by solving the equation of motion 
\begin{equation}
  \label{eq:EoM9}
  \left[ \hat{H}'_{\rm RPA}, \hat{O}^\dagger_\lambda \right] = \omega_\lambda
  \hat{O}^\dagger_\lambda
\end{equation}
for the phonon operators $\hat{O}^\dagger_\lambda$ which are linear 
combinations of the basic bosons $\hat{b}^\dagger_\mu$ and  $\hat{b}_\mu$, 
equation~(\ref{eq:boson1}).  
The linear part of the Q-operators in terms
of the boson operators $\hat b^\dagger$ and $\hat{b}$
has the following form
\begin{equation}
  \label{eq:QQpRPA2}
  \tilde{Q}_{m } = \sqrt{\varphi_m} \sum_\mu q_{m, \mu }
  \left( \hat{b}_{\mu}^\dagger + \varphi_m \hat{b}_{\mu } \right)
\end{equation}
Here $q_{m, \mu}$ are the single--particle matrix elements of
$\hat{Q}_m$ in equations~(\ref{eq:Q1}--\ref{eq:Q3}).

The equation of motion~(\ref{eq:EoM9}) leads to the following 
determinant of the secular equations (c.f.~\cite{KN86}) 
\begin{equation}
  \label{eq:QQpRPA8}
  F(\omega_{\lambda}) =  \det~({\bf R}- \frac{\bf 1}{2 \kappa}~)
\end{equation}
where the matrix elements of ${\bf R}$ are given by
\begin{equation}  \label{eq:QQpRPA6}
 R_{km}(\omega_{\lambda}) = 
 \sum_\mu \frac{q_{k,\mu }q_{m,\mu } C_\mu ^{km}}{E_\mu^2 -\omega_{\lambda}^2}
\end{equation}
involving the coefficients
\begin{equation}
  \label{eq:QQpRPA7}
  C_{\mu}^{km} = \left\{
    \begin{array}{ll}
      E_{\mu } & k = 0,-1,2 (1,-2); \quad m = 0,-1,2 (1,-2)\\ 
      \omega_{\lambda} & k = 0,-1,2 (1,-2); \quad m = 1,-2 (0,-1,2)\\ 
    \end{array}
    \right.
\end{equation}
The zeros of the function $F$ 
\begin{equation}
  \label{eq:disp0}
  F(\omega_\lambda)=0.
\end{equation}
yield the RPA eigenfrequencies $\omega_\lambda$.

Since the mean field violates the rotational invariance, among the RPA
eigenfrequencies there exist two spurious solutions.
One "spurious" solution at $\omega_{\lambda}=\Omega$ 
corresponds to a collective rotation, since 
$[H',J_{\pm}]=[H',J_x\pm iJ_y]=\mp\Omega J_{\pm}$.
The other solution with
zero frequency is associated with the rotation around the $z$ axes, since
$[H',J_z]=0$. This spurious mode allows to determine 
the Thouless-Valatin moment of inertia, 
${\cal J}_{T.V.}$~\cite{TV62,MW69,MW70},  which can be calculated 
from the equations
\begin{eqnarray}
  \label{eq:TV1}
  \left[  \hat{H}'_{\rm RPA}, i \hat{\Phi} \right] &=&
  \frac{\hat{J}_z}{{\cal J}_{T.V.}}, \\
  \label{eq:TV2}
  \left[ \hat{\Phi} , \hat{J}_z \right], &=& i. 
\end{eqnarray}
Here the angle operator $\hat{\Phi}$ is the canonical
partner of the angular momentum operator $\hat{J}_z$.

The total energy can be split in the mean field and the RPA contribution
\begin{equation}
        \label{eq:Etotal}
        E' = E'_{\rm MF} + E'_{\rm RPA}+\frac{\Omega}{2}
\end{equation}
with explicitly written the rotational "spurious" mode and 
the RPA contribution of non-spurious modes and of the spurious zero mode. 
The RPA contribution can be expressed as~\cite{Ma77,BR86}
\begin{equation}
        \label{eq:Erpa}
        E'_{\rm RPA} = \frac{1}{2} \left( \sum_\lambda \omega_\lambda - \sum_\mu
        E_\mu \right).
\end{equation}
The exchange term~\cite{Ma77,BR86} can be neglected which means to use the 
Hartree approximation.

We recall that the mean field energy, 
the quasiparticle (particle-hole) excitations, the RPA eigenfrequencies, 
are calculated in the {\it rotating frame}. In other words, 
these quantities are functions of the rotational frequency $\Omega$ 
that is our external parameter.
We remind that in the rotating frame the appropriate state 
variables are the Routhian energy  $E'$ and the rotational 
frequency $\Omega$ in contrast to 
the {\it lab} system where the appropriate state variables are 
the energy defined by the Legendre transformation $E=E'+\Omega I$ 
and the angular momentum $I$.
While the {\it operator relation} $dH'/d\Omega=-J_z$ holds, 
the corresponding relation in the RPA order
\begin{equation}
\label{qam}
dE'/d\Omega = - \langle RPA|J_z |RPA \rangle 
\end{equation}
is taken only as a reasonable approximation to the 
relation between the exact energy and the angular momentum. 
In fact, it is not only a very difficult task to calculate numerically the 
RPA expectation value $\langle RPA|J_z |RPA \rangle$.  This calculation
sensitively 
depends also from the validity of the QBA itself which raises another 
problem \cite{AF01}.   
However, the use of the total Routhian and rotational frequency 
as the appropriate variables to determine the angular 
momentum of the system is consistent with our numerical 
calculations.\\
The proper definition of the angular momentum within RPA 
is a known problem which was controversially
discussed by several authors. We mention below some important points
of these discussions.\\  
Assuming that the RPA expectation value of the angular momentum in the yrast
state is
\begin{equation}
\label{ang1}
\langle RPA|J_z |RPA \rangle=I,
\end{equation}
Reinhardt  proposed to use the equations~(\ref{qam}) and~(\ref{ang1}) to define 
the quantisation 
condition for the angular momentum (see discussion in~\cite{Re82} where 
the rotation axis is chosen to be x).
Since in the mean field approximation 
$\langle J_z\rangle= -d E'_{\rm MF}/d\Omega$ \cite{RS80,BR86}, 
we obtain from the above equations the following quantisation condition
\begin{equation}
\label{ang}
\langle J_z\rangle=I+\frac{1}{2}+\frac{d E'_{\rm RPA}}{d\Omega}
\end{equation}
According to Reinhardt, the smallness of the contribution of
the RPA modes $d E'_{\rm RPA}/d\Omega$ to the value of the angular 
momentum in the yrast state could be 
used as the validity criteria for the self-consistent cranking model.

It should be pointed out that Marshalek raised the question 
about the validity of equation (\ref{ang}) (see discussion in~\cite{Ma82}
where the rotation axis is $x$ as well).
He assumes that the mean field value of the angular momentum in the yrast state 
is $\langle J_z\rangle=I$ which  should be preserved in the RPA order too. 
He proposed to use the energy and the angular momentum as the appropriate
variables to analyse the rotational properties.
Considering the cranking Hamiltonian up to the second order of RPA, 
he proved that all conservation laws are restored 
(see Sec.4.1 in~\cite{Ma77}). Note that to prove all conservation 
laws it was important to
keep the second order of the cranking term $\Omega J_z^{(2)}$ in 
the cranking Hamiltonian as well.
From this point of view, we would find the contribution of the
second order term of $J_z$ operator to the expectation value 
$\langle RPA|J_z|RPA\rangle=\langle J_z\rangle+\langle RPA|J_z^{(2)}|RPA\rangle$
(see discussion in~\cite{EM80,N87}), which is inconsistent with
Marshalek's assumption.
Furthermore, the total energy in~\cite{Ma77} and~\cite{Ma82} 
is a sum of the mean field energy $E_{SCC}$  defined in the {\it lab} frame 
whereas the RPA correlation energy is defined in the {\it rotating} frame.
Marshalek assumed that the RPA frequencies in the {\it lab} and {\it rotating} 
frame are the same. This leads to an inconsistency 
which becomes obvious if we consider a rotation
about a symmetry axis. In this case the RPA states are characterised by 
the projection of the angular momentum upon the symmetry axis because
$[J_z,\hat{O}^\dagger_\lambda]=\lambda\hat{O}^\dagger_\lambda $
with $\lambda$ being the value of the angular momentum carried by the phonons
along the $z$ axis. We thus obtain
\begin{equation}
[ \hat{H}'_{\rm RPA}, \hat{O}^\dagger_\lambda]=
[ \hat{H}_{\rm RPA}-\Omega \hat{J}_z, \hat{O}^\dagger_\lambda]
=(\tilde{\omega}_{\lambda}-\lambda \Omega) \hat{O}^\dagger_\lambda=
\omega_{\lambda}\hat{O}^\dagger_\lambda
\end{equation}
This equation explicitly demonstrates that the RPA eigenfrequencies in 
the {\it lab} frame $\tilde{\omega}_{\lambda}$ are different from
 the RPA eigenfrequencies in the {\it rotating} frame $\omega_{\lambda}$ if
 the RPA modes carry angular momentum (see details and discussion in~\cite{bif}).

Thus, considering equations~(\ref{qam}) and~(\ref{ang1}) to be a suitable 
relation between the exact energy and the angular momentum 
and knowing the rotational dependence $E'=E'(\Omega)$ of the total energy,
it enables us to calculate
the angular momentum $I=I(\Omega)$ by the derivative
\begin{equation}
  \label{eq:Qvib}
  I = J_{\rm MF} + \Delta J_{\rm RPA} = 
  - \frac{\Delta E'_{\rm MF}}{ \Delta \Omega} 
      - \frac{\Delta E'_{\rm RPA}}{\Delta \Omega}-\frac{1}{2}
\end{equation}
and, subsequently, the dynamical moment of inertia as the second derivative 
\begin{equation}
  \label{eq:Qvib3}
  {\cal J}^{(2)} = \frac{\Delta I}{\Delta \Omega}.
\end{equation}
In practise the differentiations are numerically obtained with
finite differences $\Delta\Omega$ as indicated in the above equations. 
Hereafter, we denote the angular momentum $I$ calculated in accordance
with equation (\ref{eq:Qvib}) as $J_{\rm RPA}$.

The RPA solution $\omega_\lambda=\Omega$
that contributes to the total angular momentum by 
$-1/2$, is very sensitive to numerical
inaccuracies and to the limits of the configuration space in the
calculation. In fact, at small but nonzero rotational
frequencies the spurious solution at zero and the solution at
the rotational frequency get mixed up because of numerical
inaccuracies. The rotational solution will then no longer be exactly
at $\Omega$. Since the angular momentum is calculated as
the derivative, these errors can lead to large erratic contributions to the
angular momentum. We avoid such errors by explicitly
finding the rotational solution and subtracting from the RPA energy. 
The missing term is then  replaced by $\frac{\hbar \Omega}{2}$. 
The rotational solution can be identified by its collective nature and 
large overlap with the $\hat{J}_+\sim \hat{J}_x+i\hat{J}_y$ operator 
to which it should be identical if the solution was found exactly. 
This method is applicable when the level density is low, which is satisfied 
when $\hbar \Omega \leq 0.3$ MeV. At larger $\hbar \Omega$ the
rotational and the spurious solution decouple in a numerically
stable way and the solution at the rotational frequency is found with
a high accuracy.
To calculate the RPA energy~(\ref{eq:Erpa}) it was crucial to apply the contour
integral method developed in~\cite{DA99}.

It is important to use a large enough configuration space in the calculations
which include mixing of different $N$-shells. 
To reduce the number of 
$N$-shells needed in the calculation we always transform the matrix 
elements into the stretched basis.

\section{The RPA quadrupole correlations}
\label{volume}

In this section we focus our analysis on the contribution of 
the RPA correlations to the self-consistent mean field solution.
Parts of these results where presented in~\cite{AD01,NA02}.

The self-consistent mean field Hamiltonian agrees with that of a rotating
three-dimensionally deformed oscillator
\begin{equation}
  \label{eq:def-osc}
  \hat{H}_{MF}' = \sum_{i}{\Bigg[}\frac{p^2}{2M}+ \frac{M}{2} 
  ( \omega_x^2  x^2 + \omega_y^2  y^2 + \omega_z^2  z^2 ){\Bigg]}_i 
  - \Omega \hat{J}_z
 \end{equation}
where $\omega_{x,y,z}$ are the three oscillator frequencies~\cite{NR95}.

We start with the version (i) of the QQ-model 
($\Delta N=0, \pm 2$ shell mixing is included) 
and {\rm without} the volume conservation constraint.
The self-consistent mean field solutions of 
the Routhian, equation (\ref{eq:Routh}), are found from 
the relations
\begin{eqnarray}
  \label{eq:QQsc1}
  \kappa \left< \hat{Q}_0 \right> &=& \sqrt{\frac{5}{4\pi}}\hbar \omega_0
  \beta \cos \gamma , ~~~\nonumber\\
 \kappa \left< \hat{Q}_2 \right> &=& -\sqrt{\frac{5}{4\pi}}  \hbar
  \omega_0 \beta \sin \gamma ,~~~\\
 \left< \hat{Q}_{\pm 1, -2} \right> &=& 0\nonumber 
\end{eqnarray}
where, as usual, the contributions of exchange terms are omitted.

Two systems  are considered: (a) the lighter nucleus N=Z=14 
with the self--consistent deformation $(\varepsilon_2=0.3, \gamma=0,$ 
at $\Omega=0 )$ 
and (b) the nucleus N=Z=76  defined with  the corresponding set 
$(\varepsilon_2=0.3, \gamma=0,$ at  $\Omega=0 )$  
by choosing the appropriate strength $\kappa$ for each system.
The largest $N$-shell included in the calculation is $N=6$ in (a) 
and $N=8$ in (b).

\begin{figure}[htbp]
  \centerline{\includegraphics[clip,height=7.5cm,angle=-90]{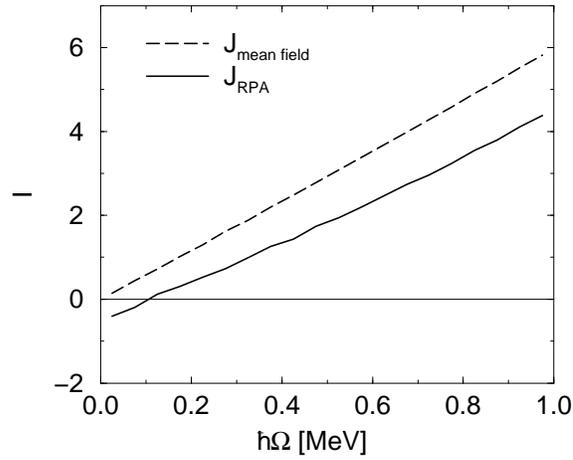}}
  \caption{The angular momentum, in units of
    $\hbar$, for a rotating deformed harmonic oscillator with
    (solid line) and without (dashed line) QQ ($\Delta N=2$) isoscalar
    RPA correlations as a
    function of $\Omega$. The self-consistent deformed oscillator is
    filled with 14  protons and 14  neutrons.}
  \label{fig:Eom2a14}
\end{figure}

\begin{figure}[htbp]
  \centerline{\includegraphics[clip,height=7.5cm,angle=-90]{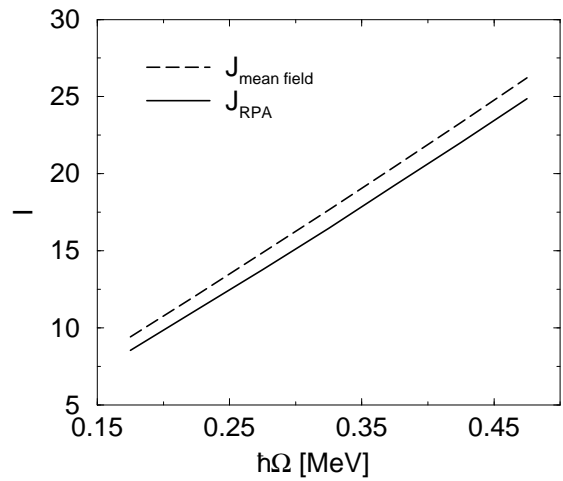}}
  \caption{The angular momentum, in units of
    $\hbar$, for a rotating deformed harmonic oscillator with
    (solid line) and without (dashed line) QQ ($\Delta N=2$) isoscalar
    RPA correlations as a
    function of $\Omega$. The self-consistent deformed oscillator is
    filled with 76 protons and 76 neutrons.}
  \label{fig:Eom2a76}
\end{figure}

It will be seen that both systems behave qualitatively 
similar with respect to the rotation. 
 By searching for the energy minimum
for finite rotational frequencies the self-consistent mean field solutions
are found. This minimisation 
makes also sure that the RPA will restore
the rotational invariance.
The mean field angular momentum for these  systems  calculated by means 
of equation~(\ref{eq:Qvib}) is shown at 
Figs.~\ref{fig:Eom2a14}, ~\ref{fig:Eom2a76}. It is 
proportional to the rotational frequency $\Omega$.
This means that our system even if the deformation is
changing with the rotation gets an approximately constant moment of
inertia~\cite{Ra87}.

Taking into account the RPA
correlations, we obtain a substantial decrease of the
total energy $E'$ and also a dependence of 
the RPA angular momentum $J_{\rm RPA}$ on $\Omega$. It turns
out that $J_{\rm RPA} \propto \Omega$.  
The dynamical moment of inertia is approximately constant. 
However, it is reduced 
(the reduction is 16\% and 4\% in N=Z=14 and 76, respectively) 
due to the RPA correlations, as seen in figure~\ref{fig:Eom2a14}.
The RPA correlations cause relatively smaller effect on
the moment of inertia in the heavier system  than in the lighter one.
This is because of the absolute $\Omega$-dependence of 
the $E_{\rm RPA}$ becomes slightly smaller and  the mean field moment 
of inertia becomes much larger in the heavier system.
The spherical oscillator part in our Hamiltonian, 
equation (\ref{eq:QQ1}), can in these cases be 
replaced by any set of spherical single-particle energies. This does
not affect the general properties of the RPA correlations. 
The result would not qualitatively
change when including a spin-orbit splitting.

Now the above systems are studied for the version (ii), i.e. by dropping the
$\Delta N=\pm2$ matrix elements of the quadrupole operators $\hat{Q}_m$,
equations~(\ref{eq:Q1})-(\ref{eq:Q3}). We recall that the  quadrupole
Hamiltonian with $\Delta N=0$ is  commonly used 
for obtaining the self-consistent solutions which includes both non-zero
spin-orbit terms as well as the QQ-forces~\cite{RS80}. 
This task needs much less efforts, since solutions can be found in a relatively
small configuration space. Having in mind that the $\Delta N=2$
particle-hole excitations are energetically unfavored relative 
to the $\Delta N=0$,
one would expect a reliable results within the model.
In addition, the effect of the higher lying $\Delta N=2$ vibrations 
could be studied. 
The self-consistent mean field solutions of 
the Routhian  in equation~(\ref{eq:Routh}) are again found from 
the relations~(\ref{eq:QQsc1}) where this time the
quadrupole operators are written in the stretched 
representation~\cite{Ni55,NR95}.

\begin{figure}[htbp]
  \centerline{\includegraphics[clip,height=7.5cm,angle=-90]{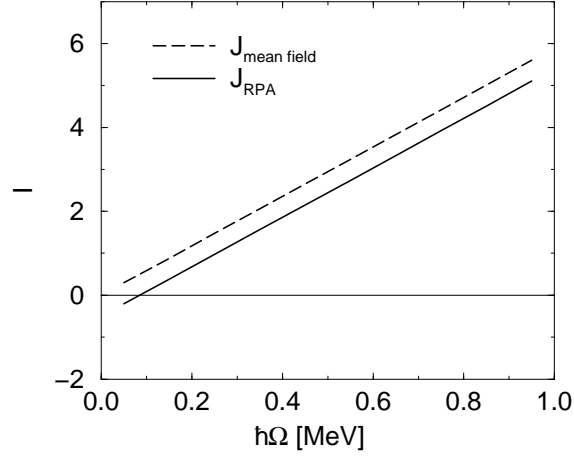} }
  \caption{The angular momentum, in units of
    $\hbar$, for a rotating deformed harmonic oscillator with
    (solid line) and without (dashed line) QQ ($\Delta
    N=0$) isoscalar RPA correlations as a
    function of $\Omega$. The self-consistent deformed oscillator is
    filled with 14 protons and 14 neutrons.}
  \label{fig:Eom2b}
\end{figure}

In this model the angular momentum is as above proportional to $\Omega$
but the RPA correlation energy does not any more influence the moment
of inertia as can be seen in figure~\ref{fig:Eom2b}.
We conclude that an important part of the RPA correlations originate from
the $\Delta N=2$ part of the QQ-forces. 
It is interesting to mention that the simple QQ-forces with
$\Delta N=0$  provide quite reasonable results with regard equilibrium 
deformations in the mean field calculations.

An explicit calculation of the commutation relation
\begin{equation}
  \label{eq:com1}
  \left[ \hat{H}'_{\rm RPA}, \hat{J}_z \right] = 0 , 
\end{equation}
gives a stringent check of the RPA restoration of the rotational
symmetry. In the numerical calculation of the N=Z=14 case above we
found that equation~(\ref{eq:com1}) was fulfilled with an accuracy of
$10^{-3}$ when the deformation parameters $\varepsilon_2$ and $\gamma$ were
determined with the accuracy of 4 and 2 decimal figures, respectively.

Next, we apply the volume conservation constraint and the double-stretched
representation of quadrupole operators and
compare the results with those obtained with the (normal) QQ-forces.
We remind that the double-stretched representation is based on the condition 
such that the change in the density distribution must be accompanied 
with a change in the potential~\cite{BM75}.

For the triaxially deformed oscillator
 the self-consistency condition \cite{AT}
 \begin{equation}
  \label{eq:dsQQ5}
  \omega_x^2 \left< x^2 \right> =\omega_y^2 \left< y^2 \right>
  =\omega_z^2 \left< z^2 \right>
\end{equation}
 and the volume conservation constraint $\omega_x\omega_y\omega_z=\omega_0^3$
yield a self-consistent residual quadrupole interaction ~\cite{Ab85,SK89}
\begin{equation}
  \label{eq:dsQQ1}
  \hat{H}' = \hat{H}_{MF}' - \hat{H}_{\rm res}
\end{equation}
where $\hat{H}_{MF}'$ is defined by equation~(\ref{eq:def-osc}) and 
\begin{equation}
  \label{eq:dsQQ2}
  \hat{H}_{\rm res} = \frac{\overline{\overline{\kappa}}}{2} 
  \sum_{m=0,\pm 1,\pm 2}
  \overline{\overline{Q}}_m \overline{\overline{Q}}_m .
\end{equation}
Here $\overline{\overline{Q}}$ is the quadrupole operator written in doubly
stretched coordinates
\begin{equation}
  \label{eq:dsQQ3}
  \overline{\overline{x}}_i = x_i \frac{\omega_i}{\omega_0} ,\qquad i=1,2,3 ,
\end{equation}
where $\omega_i$ is one of the oscillator frequencies. The self-consistent
$\overline{\overline{\kappa}}$ can be found as~\cite{SK89}
\begin{equation}
  \label{eq:dsQQ4}
  \overline{\overline{\kappa}} = \frac{4\pi}{5}
  \frac{M \omega_0^2}{A \left< \overline{\overline{r}}^2 \right>}.
\end{equation}
From the self-consistency condition, equation~(\ref{eq:dsQQ5}), follows that
\begin{equation}
  \label{eq:dsQQ6}
  \left<\overline{\overline{Q}}_m \right> =0 ,\qquad m=0,\pm 1,\pm 2.
\end{equation}
The $\overline{\overline{Q}}$ operators do therefore not affect the mean field
energy and can be used directly in the RPA formalism.

\begin{figure}[htbp]
  \centerline{\includegraphics[clip,height=7.5cm,angle=-90]{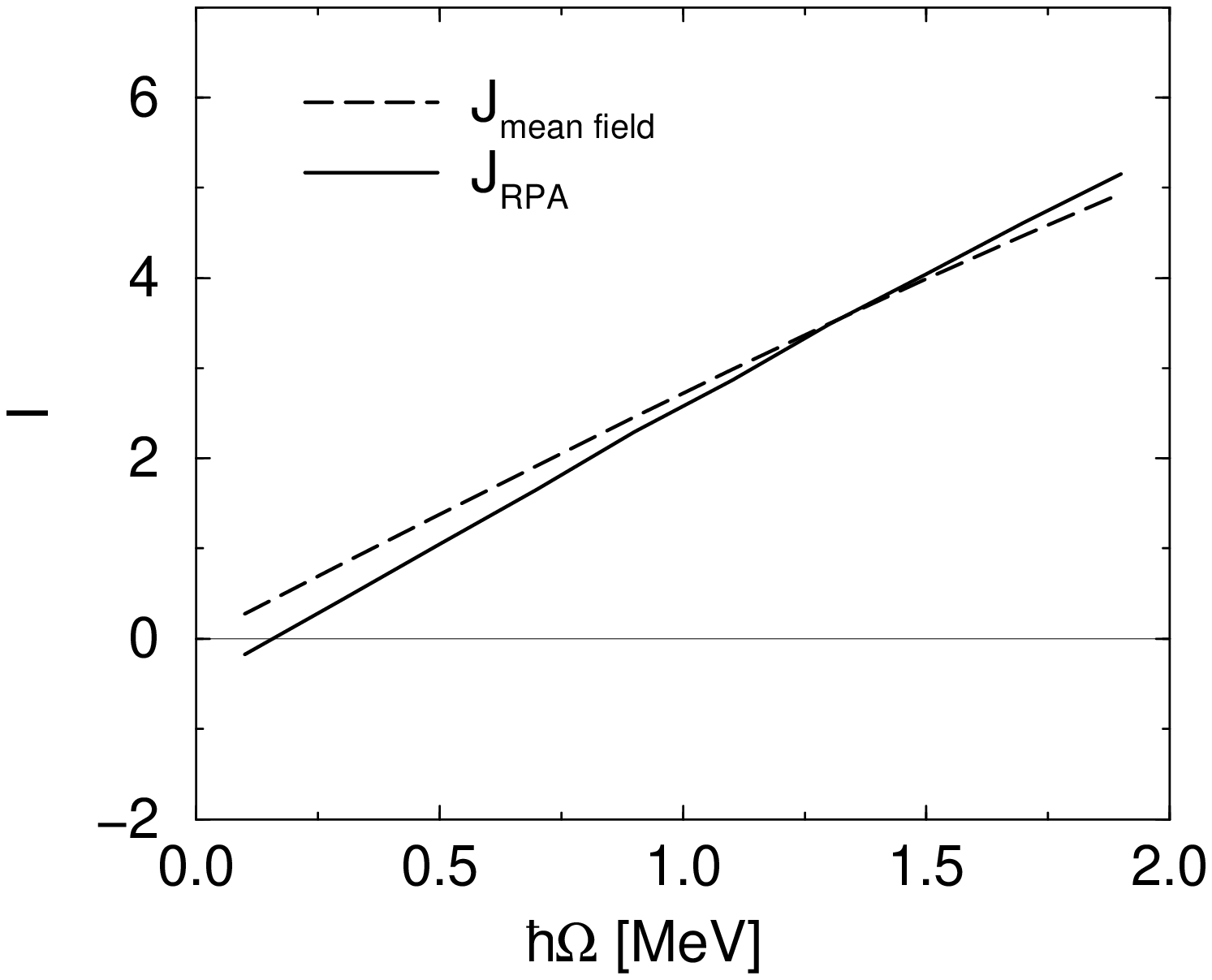}}
  \centerline{\includegraphics[clip,height=7.5cm,angle=-90]{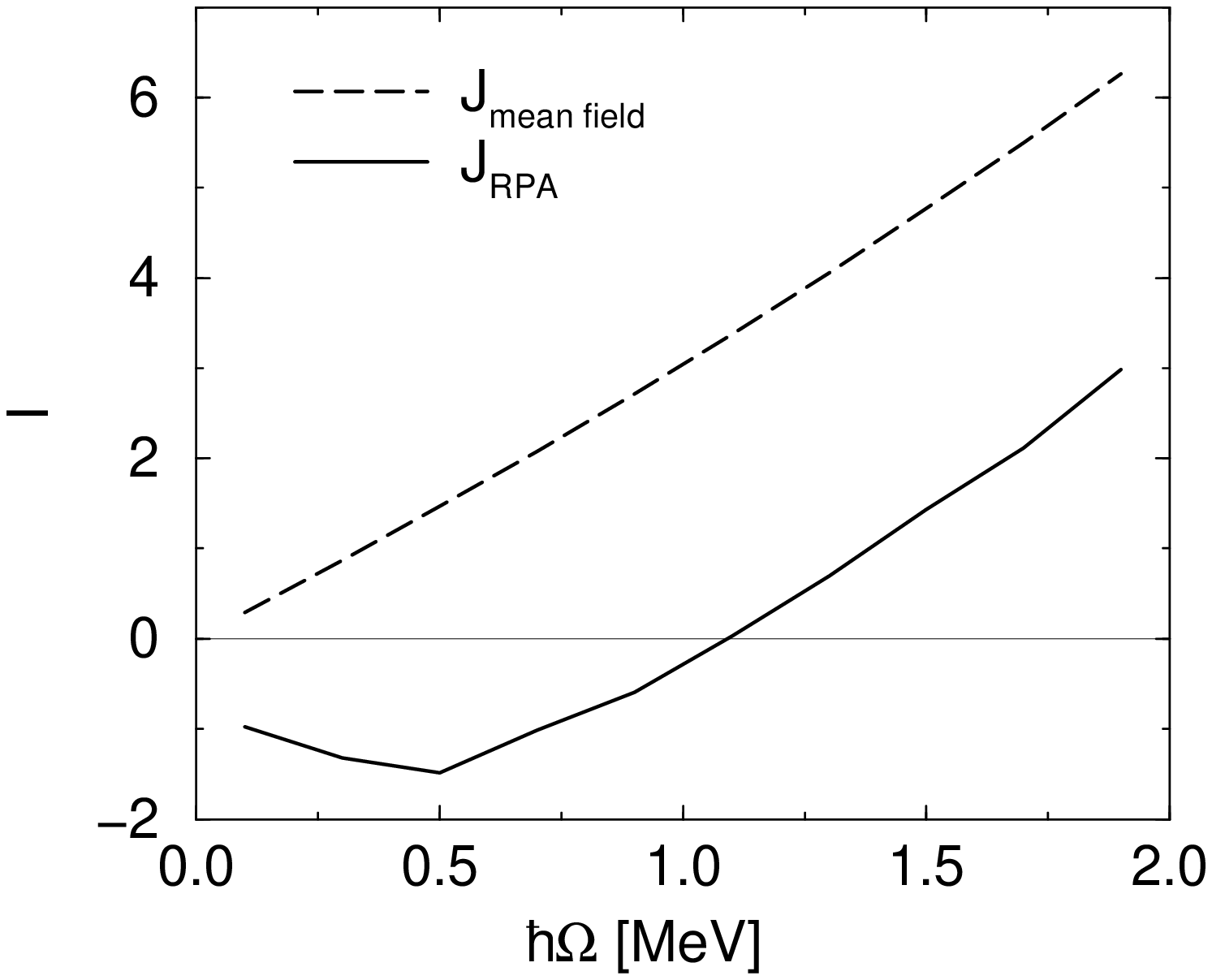}}
  \caption{The angular momentum, in units of $\hbar$, for a rotating
    deformed harmonic oscillator with
    (solid line) and without (dashed line) isoscalar RPA correlations
    as a function of $\Omega$. Upper part: volume conservation, lower
    part: QQ-forces.
    The self-consistent deformed oscillator is
    filled with 10 protons and 10 neutrons. Both systems have the same
    deformation at $\Omega=0$.}
  \label{fig:HO10}
\end{figure}
\begin{figure}[htbp]
  \centerline{\includegraphics[clip,height=7.5cm,angle=-90]{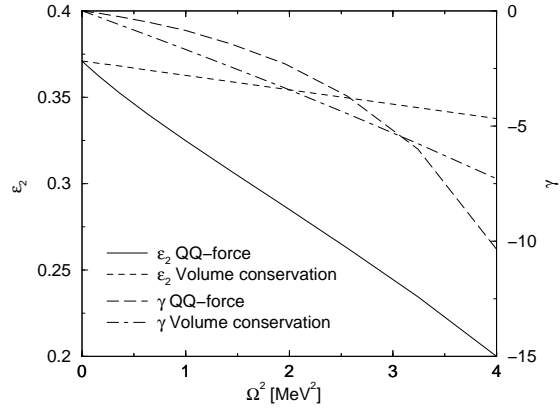}}
  \caption{The deformation parameters $\varepsilon_2$ and $\gamma$ 
        of a rotating deformed harmonic oscillator calculated with the volume 
        conservation constraint and with the QQ-forces. 
        The self-consistent deformed oscillator is filled with 
        10 protons and 10 neutrons.}
  \label{fig:betagamma}
\end{figure}

In figure~\ref{fig:HO10} results for the harmonic oscillator with 10
protons and 10 neutrons  are presented.
As for the cases studied above the QQ forces give a
reduced angular momentum and lead to a delayed start of the
rotational band at $\hbar \Omega=1.1$ MeV (where the calculated angular 
momentum becomes positive). The RPA correlations obtained with the 
use of the double-stretched residual interaction
results in an increased dynamical moment of inertia in 
contrast to the results with the QQ-forces.
The rotational dependence of the equilibrium deformations is presented
in figure~\ref{fig:betagamma}.   
The minimum calculated with the volume conservation constraint 
seems to be changing more slowly then with the QQ-forces.
\begin{figure}[htbp]
  \centerline{\includegraphics[clip,height=7.5cm]{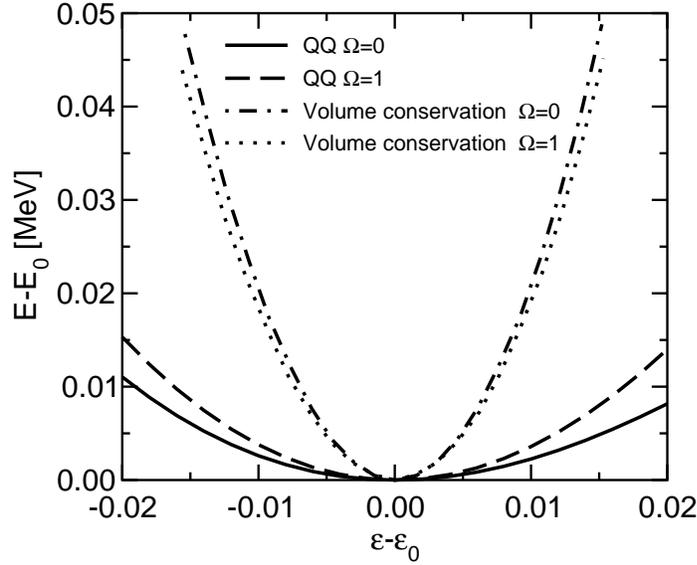}}
  \caption{Mean field energy as a function of deformation for the
    harmonic oscillator using the QQ-forces and the volume
    conservation constraint. For each method the curve is plotted for the
    non-rotating system and for $\hbar \Omega=1$ MeV.}
  \label{fig:HOdef10}
\end{figure}

 From figure~\ref{fig:HOdef10}, which shows the mean field energy as a
function of the deformation at different rotational frequencies with 
and without the volume conservation constraint, one can qualitatively 
understand the different results. The RPA contribution to the angular 
momentum is obtained as the negative derivative of the correlation energy.
The QQ-forces give a very flat minimum that is changing relatively fast
with the rotational frequency. This causes a large correlation energy that
also is changing relatively fast with the rotational frequency.  As a
consequence, there is the large contribution from the RPA correlations 
to the angular momentum seen in figure~\ref{fig:HO10}. A small contribution 
to the total angular momentum could be explained as the result 
of the increase of the rigidity of the potential with the increase of 
the rotational frequency $\Omega$. The flat potential energy surface 
is also the reason why we obtain a delayed start 
of the rotational band using the QQ-forces. 
With the volume conservation constraint one observes an opposite
behavior. The minimum is much more distinct and is less affected by
the rotation. In figure~\ref{fig:HO10} a small increase in the dynamical
moment of inertia can be seen which is consistent with the fact that
the potential gets flatter with increasing $\Omega$ when using the volume
conservation constraint.

\begin{figure}[htbp]
  \centerline{\includegraphics[clip,height=7.5cm,angle=-90]{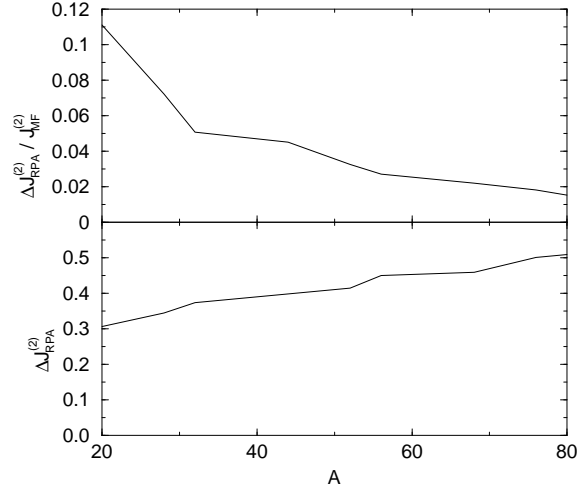}}
  \caption{RPA contribution to the moment of inertia as a function
        of mass number A for N=Z harmonic oscillator with 
        the volume conservation constraint.
        The plotted values are for $\Omega=0$ (MeV).
        Upper panel: relative contribution. Lower panel: absolute contribution.}
  \label{fig:MoIA}
\end{figure}
In figure~\ref{fig:MoIA} the mass dependence of the RPA contribution to the
moment of inertia, as calculated with the volume conservation constraint and 
the double-stretched quadrupole residual interaction, is plotted. Even though 
the RPA contribution to ${\cal J}^{(2)}$
is increasing with increasing a mass number A 
the relative contribution is decreasing. This is
consistent with the simple picture that the mean field approximation 
works better in heavier systems.

\section{Summary}
\label{summary}

The value of the angular momentum and the moment of inertia are
generally dependent on the correlations induced by the shape
vibrations. The size of the RPA correlations depends on the curvature
of the mean field potential. Their influence on the angular
momentum and on the moment of inertia is determined
on how the potential curvature depends on the rotational frequency. This
dependence is different in different mass regions and is determined by 
the degree of filling  of the shell. 
The scale of the variation of the correlations is sensitive to
the accuracy of the mean field approximation and to the size of 
the configuration space.

The very flat potential of the QQ-forces leads to 
large effects of the quadrupole correlations  
upon the angular momentum and the moment of inertia. 
In fact, the large RPA contributions for the QQ-forces 
cause the delay of the rotation in contrast to the results obtained with 
the volume conservation constraint. In addition, the QQ-forces, 
without the volume conservation condition, may not produce any stable minimum
above a certain deformation and it seems that the
deformation in the considered cases is close to that limit. When
comparing the results with and without the $\Delta N =2$ 
mixing one notices that even when the two calculations provide 
similar mean field solutions, which is expected~\cite{Ni55}, the
RPA does yield significant differences. 
We conclude that these forces are not very
realistic in describing RPA correlations in nuclear structure.

The RPA calculations with the volume
conservation constraint and the double-stretched quadrupole 
residual interaction give a smaller contribution to the angular momentum. 
The RPA contributions to the moment of inertia is almost constant 
as a function of the rotational frequency but
show a strong dependence on the mass number. 
Even though the absolute contribution
is increasing with increasing the mass number, 
the relative contribution is decreasing from 11\% at A=20 to less then 2\% 
at A=80. The RPA based on the double-stretched 
quadrupole residual interaction provides a more realistic
result in contrast to the cases of the QQ-forces discussed above.

Finally, we conclude that the volume conservation constraint is very important
requirement when considering effective forces. 
Further analysis of the self-consistency 
between a mean field approximation and a treatment of the
quadrupole shape vibrations for various realistic forces is needed 
to make reliable comparisons with experimental results. Another important 
question which has to be
addressed in further investigations is how other multipoles such as
monopole and octupole ones would affect the present results.

\ack
This work was supported by the UK Engineering and Physical Sciences Research 
Council (EPSRC) under grant GR/N15672 and  by Grant No.\ BFM2002-03241 
from DGI (Spain). R. G. N. gratefully acknowledges support from the 
Ram\'on y Cajal programme (Spain).

\end{document}